\documentclass[english,twocolumn]{revtex4}

\usepackage[T1]{fontenc}
\usepackage[latin9]{inputenc}
\usepackage{enumitem}
\usepackage{graphicx}
\usepackage{verbatim}
\usepackage{amsmath,amssymb,amstext,amsthm,bbm,mathtools}

\makeatletter

\usepackage{booktabs}
\usepackage{braket}
\usepackage{tikz}

\makeatother

\usepackage{babel}

\begin{document}

\title{Minimally complex ion traps as modules for quantum communication and computing}
\date{\today}
\author{Ramil Nigmatullin$^1$}
\author{Christopher J. Ballance$^2$}
\author{Niel de Beaudrap$^3$}
\author{Simon C. Benjamin$^1$}
\affiliation{$^1$Department of Materials, $^2$Department of Physics, $^3$Department of Computer Science,\\
University of Oxford, Parks Road, Oxford, UK.}
\begin{abstract}
Optically linked ion traps are promising as components of network-based quantum technologies, including communication systems and modular computers. Experimental results achieved to date indicate that the fidelity of operations within each ion trap module will be far higher than the fidelity of operations involving the links; fortunately internal storage and processing can effectively upgrade the links through the process of purification. Here we perform the most detailed analysis to date on this purification task, using a protocol which is balanced to maximise fidelity while minimising the device complexity and the time cost of the process. Moreover we `compile down' the quantum circuit to device-level operations including cooling and shutting events. We find that a linear trap with only five ions (two of one species, three of another) can support our protocol while incorporating desirable features such as {\em global control}, i.e. laser control pulses need only target an entire zone rather than differentiating one ion from its neighbour. To evaluate the capabilities of such a module we consider its use both as a universal communications node for quantum key distribution, and as the basic repeating unit of a quantum computer. For the latter case we evaluate the threshold for fault tolerant quantum computing using the surface code, finding acceptable fidelities for the `raw' entangling link as low as $83\%$ (or under $75\%$ if an additional ion is available).
\end{abstract}
\maketitle

\section{Introduction}

In order to realise the promise of quantum technologies it is highly desirable to create modular units that can interlink optically, each having an  internal storage and processing capacity. Long range quantum communication networks will require repeaters to overcome photon loss and accumulated noise~\cite{LiangmultiGenerationsOfRepeater}; generally a small quantum processor with optical outputs can be seen as a universal communications node suitable for supporting any network-based task. Meanwhile, for quantum computing the modular approach could be used to build a large-scale machine on a single site~\cite{MonroeLargeScaleModular,NemotoDiamondArchiecture,PhysRevX.4.041041}. Here modules may be called `remote' but they might be separated by centimetres or less. Several technologies have shown the in-principle capability to serve as photonically interlinked modular cells, including ion traps~\cite{Hucul2015}, nitrogen-vacancy centres in diamond~\cite{Pfaff532,Hanson2016MemoryVersusrepeatedEnt}, and superconducting qubits~\cite{PhysRevLett.112.170501,Devoret2016superconductingEnt}. Here we focus on strategies for exploiting ion traps. However much of what follows, including the purification circuits that we analyse, can be equally useful in diamond or superconducting approaches. All systems of this general kind will have the same desiderata: high purifying power with low time cost and low system complexity.

Trapped ion systems are one of the most mature quantum technologies. A variety of trap devices now exist, but in all cases electromagnetic fields are configured to spatially confine an ion plasma. In high vacuum and under the action of laser cooling, the ions organise into Coulomb crystals. The electronic and spin states of the ions can be manipulated using optical and microwave techniques, and thus each ion can embody a qubit once a suitable pair levels are identified. Coupling between the ions due to the vibrational mode of the Coulomb crystal implies the possibility of controlled manipulation of quantum states of two-or-more ions -- multi-qubit gate operations. High fidelity quantum operations in ion traps have been demonstrated by a number of groups worldwide \cite{Benhelm2008,Ballance2015a,Harty2014,Wineland2016}. Proof of principle experiments have demonstrated several quantum algorithms in single crystal devices; ranging from Shor's algorithm~\cite{Monz1068} to simulation of quantum Ising spin chains~\cite{Richerme2014,Britton2012}. 

Spectral crowding means that the larger the ion crystal the more difficult it is to coherently control the quantum dynamics of individual ions. For this reason, it is advantageous to adopt a modular design: decomposition of a device into a large number of interconnected ion traps. The links between the traps can be implemented either by shuttling of the ions between the traps or with a hybrid system, for example, with photonic interfaces. The advantage of the hybrid ion/photon approach is that it does not involve the design and manufacture of large traps with complex electrode geometries. The disadvantage is that, at the moment, the ion/photon interface is significantly noisier and far slower than the operations within an isolated ion crystal~\cite{Hucul2015}. Protocols that mitigate this network noise, such as entanglement purification, inevitably add additional resource cost~\cite{Nickerson2013,PhysRevX.4.041041}. Surprisingly it has been shown that the resource cost associated with the adoption of the flexible network architecture varies little over a wide range of module sizes \cite{Li2015}. 

The focus of this paper is to design the simplest possible ion trap modules, suitable for optically linking into a scalable communications network or a single-site modular quantum computer~\cite{Monroe2016codesign}. In our analysis, we assume gate fidelities that are already accessible in state of the art experiments. In Section~\ref{sec:trapAndInterface} we note the general requirements for optically linking modules. In Section~\ref{sec:designConsiderations}, we list our design priorities and describe the extent to which they will be met. Section~\ref{sec:PurProtocol} then provides a systematic construction of the purification protocol and shows that using three ions can reduce the infidelity of a Bell pair from $\epsilon$ to $\frac{2}{9}\epsilon^2+O(\epsilon^3)$. A suitable device structure and a device-level specification is given in Section~\ref{sec:layout}. In Section~\ref{sec:twoConnectedMods}, we numerically evaluate the performance of the node -- the final fidelity and the running time of the protocol as a function of network fidelity. In Sections~\ref{sec:communication} and \ref{sec:computing}, we indicate the performance in two practical applications: communication and fault tolerant computing. For the latter, we evaluate the fault tolerance threshold for a toric code. Finally we offer some conclusions in Section~\ref{sec:conclusions}.

\begin{figure}
\centering
\includegraphics[scale=.62]{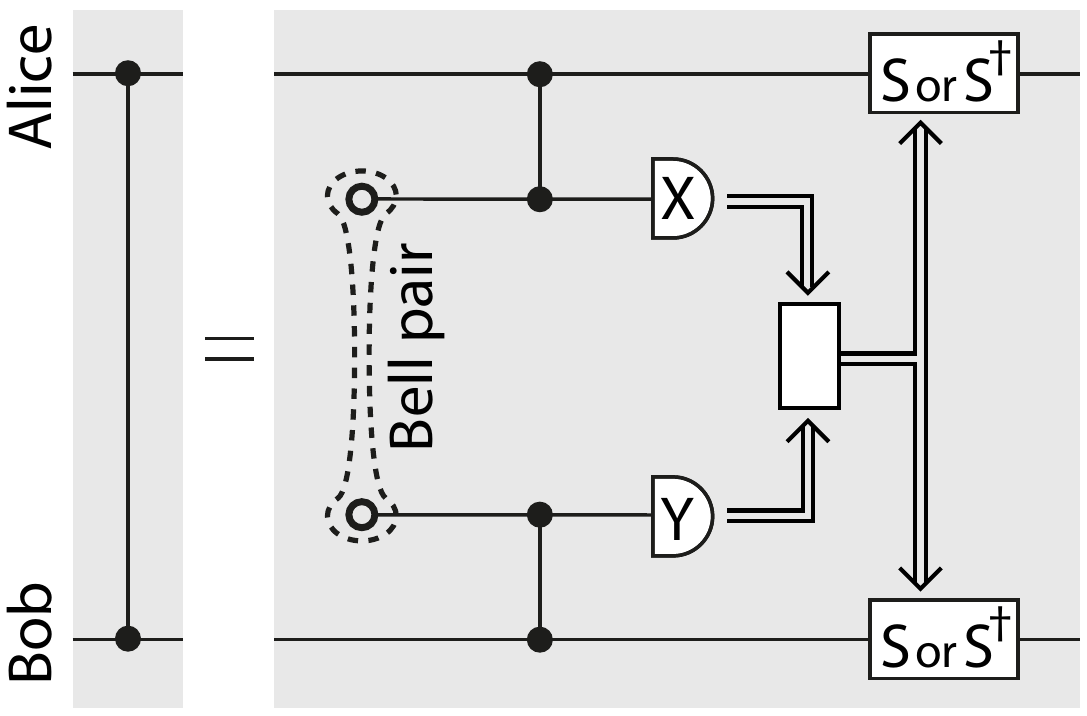}
\caption{The basic principle of performing a remote control-phase gate (left) by sharing a Bell pair and performing local operations plus classical communication (right).
\label{fig:gateTeleport}}
\end{figure}

\section{Interlinking Ion traps via photons \label{sec:trapAndInterface}}

In any modular quantum technology it will be necessary to achieve entanglement that spans the modules. For computing applications, we will wish to be able to perform quantum gates between qubits in remote locations. An elegant and practical route to achieving this is to create a shared Bell pair between  the two modules (through a process that might be probabilistic, provided that success is heralded) and then consume this Bell pair in order to implement the gate. Suppose that Alice and Bob each have a module, and within each module is a single ``application qubit'', i.e. a qubit that is part of the overall task that Alice and Bob are performing, be it communication or computing. Suppose that they also share a Bell pair $\ket{\Psi^+}\equiv(\ket{00}+{11})/\sqrt{2}$. 

If Alice and Bob would like to perform a control-phase (cPhase) gate operation between their application qubits, then they can do by the process shown in Fig.~\ref{fig:gateTeleport} which involves three steps:  (1) They each perform local cPhase operations between their application qubit and their `half' of the Bell pair,
(2) Alice measures her Bell qubit in the $x$-basis while Bob measures his in the $y$-basis, and finally (3) they each apply a single qubit gate to their application qubit. The required gate depends on the measurement outcomes at the second stage; if their measurements were the same (both measured in the positive direction on their apparatus, or both negative) then Alice and Bob should both apply single qubit gate $S\equiv{\rm diag}\{1,i\}$, otherwise they should apply $S^\dagger$. This is a simple instance of gate teleportation and the process can be verified in a few lines, see Appendix 1. Adopting  this approach to remote gate operations, the challenge of realising a modular quantum computer becomes ``How can we create {\em high fidelity} shared Bell pairs?''. 

The demand for high fidelity is crucial since any imperfection on the Bell pair will translate to noise on the remote gate operation. It is a reasonable presumption that the entanglement channel creates `raw' Bell pairs with a fidelity far below that of the local gates. In ion trap experiments, all local gates have been demonstrated with fidelities of $99.9\%$ or higher~\cite{Ballance2015a,Wineland2016}, while entanglement between traps has been achieved at the level of about $85\%$~\cite{Hucul2015}. While we can expect these numbers to continually improve, it may be that local operations are always of superior fidelity to the entangling channel; whenever this is the case, we may wish to perform purification in order to effectively upgrade the channel fidelity to a level comparable to the local gates.

In the following, we will refer to the ancillas which represent the purified Bell pairs as the ``envoy'' qubits; their role in mediating the link between modules is ``high status'' in the sense that they embody the superior, purified entanglement and they interact directly with the crucial ``application qubits''. Below the envoys are other ancilla qubits which process the lower grade entanglement, or create the `raw' entanglement between modules. However these lower status ancillas will never interact directly with the application qubit. Thus, an ion trap module will contain several ions with different designated functions: raw entanglement ion, purification ions, the envoy ion and the application ion. 

We will not discuss the means by which raw entanglement is created, except that we assume it is done optically and that the fidelity achieved is relatively poor. The details of the process are of course important since they determine the nature of the infidelity on the raw pair. Here we will assume a noise model where all imperfections are equally likely. In reality a given method for Bell state generation (e.g. Ref.~\cite{Hucul2015}) will have a unique noise spectrum; generally however, structure in the noise will make it easier to purify, and therefore our assumption of structureless noise means that the performance metrics we predict tend to the conservative end.

\begin{figure*}
\centering
\includegraphics[scale=0.4]{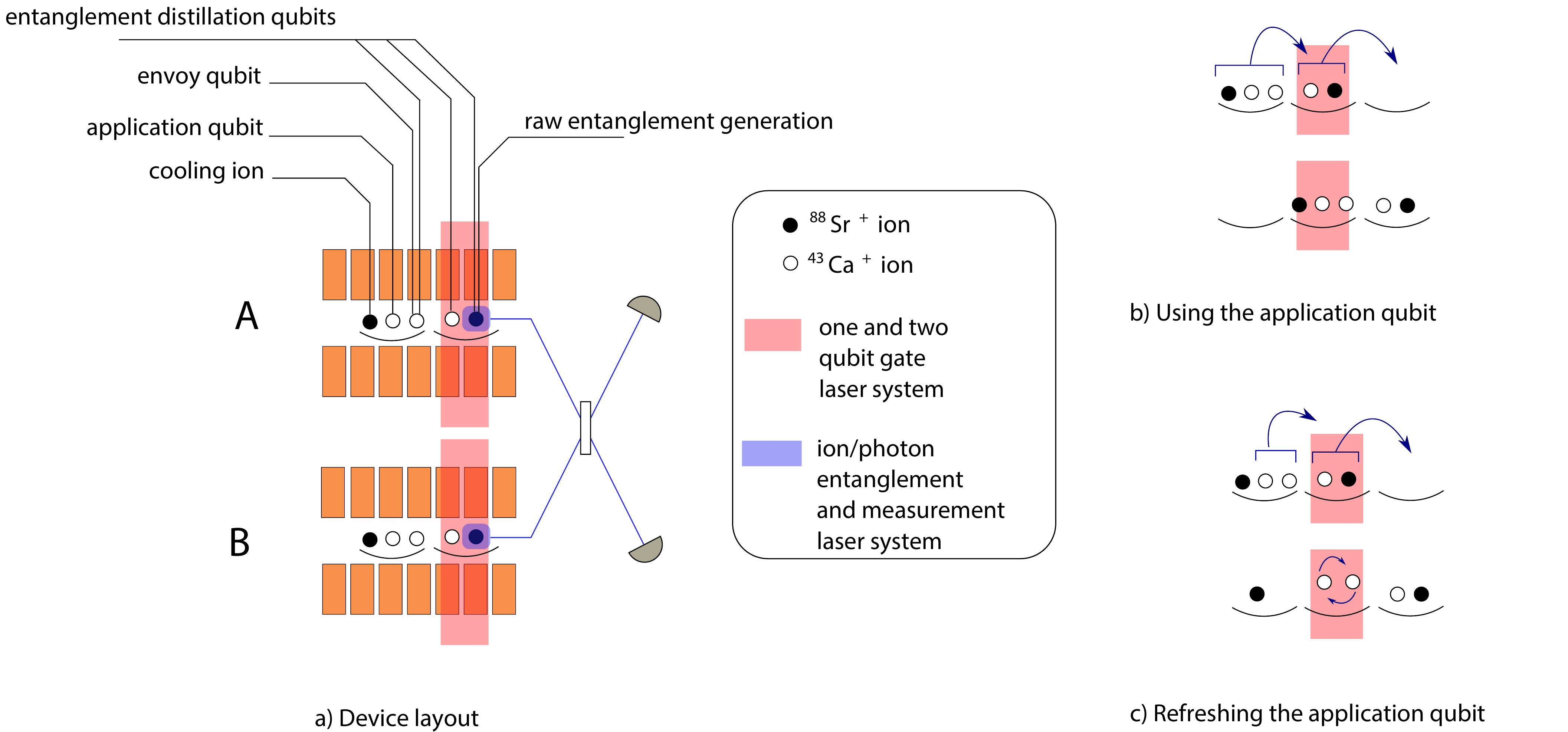}
\caption{Diagram illustrating the basic units of a proposed ion trap quantum
network. Entanglement is generated between the two nodes $A$ and
$B$ via a noisy photonic link. A purification protocol consisting
of single qubit gates, two qubit gates and measurements, generates
high fidelity entangled envoy qubits from raw noisy entangled qubits.
\label{fig:nodes_illustration}}
\end{figure*}

\section{Design considerations \label{sec:designConsiderations}}

Following the reasoning in the previous section, we now proceed to design the minimally complex ion trap that suffices as a module for communications or computing over an imperfect network. We assume that each trap contains {\em only one} ``application qubit'' and that all other ions exist only in order that the sole application qubit can perform high fidelity gates with partners in other traps. Generalisations to variants with two or more application qubits per trap are straightforward, but by focusing on this minimal device we can address the question ``What is the simplest ion trap that can suffice as a module of a quantum technology?''.

 The following features are desirable for a practical module:
\begin{enumerate}[leftmargin=*]
\item A high level of purification should be achieved (at least an order of magnitude reduction to infidelity).
\item The time cost of the purification should be modest.
\item The trap geometry should be simple, ideally linear.
\item The trap should have as few zones as possible. \label{zoneItem}
\item At most two ion species should be used (ideally one).
\item Shutting/permuting  of ions should be minimised.
\item Two-qubit gates are preferable to higher order gates. \label{twoQubitGateItem}
\item Two-qubit gates should involve neighbouring ions. \label{neighboursItem}
\item The fewest possible measurement/control systems (lasers, lenses, detectors etc) should be required.
\item The issue of cross talk, e.g. unwanted interaction of a laser or emitted photons with another ion, should be minimised by design. \label{crossTalkItem}
\item The need for sympathetic cooling must be allowed for.
\end{enumerate}

These desiderata are largely self-explanatory. The need for a purifying factor of at least ten follows from the fact that we can expect the `raw' entanglement fidelity to be at least ten times worse than the local gate fidelity (in present experiments it is two orders worse). The need to minimise the time cost follows from the fact that it will be challenging to create entanglement before significant decoherence has occurred to the application qubit. By `zone' in point (\ref{zoneItem}) we refer to a region of the trap that is significantly remote from other regions, effectively forming  a sub-trap; one zone may have several electrodes to define it and move ions. Point (\ref{twoQubitGateItem}) is motivated by the observation that experiments to date have reported lower fidelity as the number of ions involved in a gate increases, while point (\ref{neighboursItem}) results from the observation that, while a two-qubit gate is possible with a passive ion in-between, this is non-optimal and becomes more difficult with more intervening passive ions.

We find that by permitting ourselves two species of ion, the other desiderata can be satisfied to a remarkable degree. For the variant that we analyse in greatest detail,

\begin{enumerate}[leftmargin=*]
\item 10\% raw infidelity is purified to $0.6\%$ infidelity.
\item Average time cost is a factor of $\sim 8$. 
\item A linear geometry does suffice.
\item Only two zones are required.
\item Two species are employed, e.g. Ca and Sr.
\item Only one ion performs any shuttling, and ions need never be permuted$^*$. \label{permuteItem}
\item Only one-qubit and two-qubit gates are employed.
\item All two-qubit gates are on nearest-neighbours.
\item Only a single instance of each control/measurement system is required for each entire trap device.
\item By adopting a {\em global control} principle, laser cross talk is negligible; laser beams need not be tightly focused.
\item Cooling is efficiently integrated via a dual-role ion.
\end{enumerate}

The asterisk in point  (\ref{permuteItem}) is present because for certain functions, such as fault tolerant surface code computing, it may be desirable to periodically exchange the roles of application qubit and the envoy.  An efficient way to do this would be to physically permute the two ions so that they exchange places; however if this is not possible, a logical SWAP operation will suffice instead. The reason that this exchange may be desirable is explained presently when we appraise the module's performance for fault tolerant computing.

Generally the desired features are quite compatible with one another, and in particular it is quite natural to support (\ref{twoQubitGateItem}) and (\ref{neighboursItem}) because of the tiered nature of the purification process. 
The use of two zones proves to be very valuable in meeting the other desiderata especially point (\ref{crossTalkItem}). In the following we will use the label `rowdy end' for the zone in which raw entanglement is created and all measurements are performed. The term `tranquil end' will be used for the zone where the application qubit resides. The envoy qubit will shuttle between the two zones, ultimately delivering purified entanglement to the application qubit in order to perform the remote gate.  Thus at any one time each module would contain only two crystals with 2-3 ions.

\section{Purification Protocol \label{sec:PurProtocol}}
In this section, we specify the purification circuit that we have simultaneously designed along with the ion trap layout described in the following section. We will use these primitives: generation of raw entangled pairs, two qubit gates between adjacent qubits, and measurements. Presently we will `compile down' the circuit to a set of device-level operations including cooling and shuttling operations.

\begin{figure*}
\centering
\includegraphics{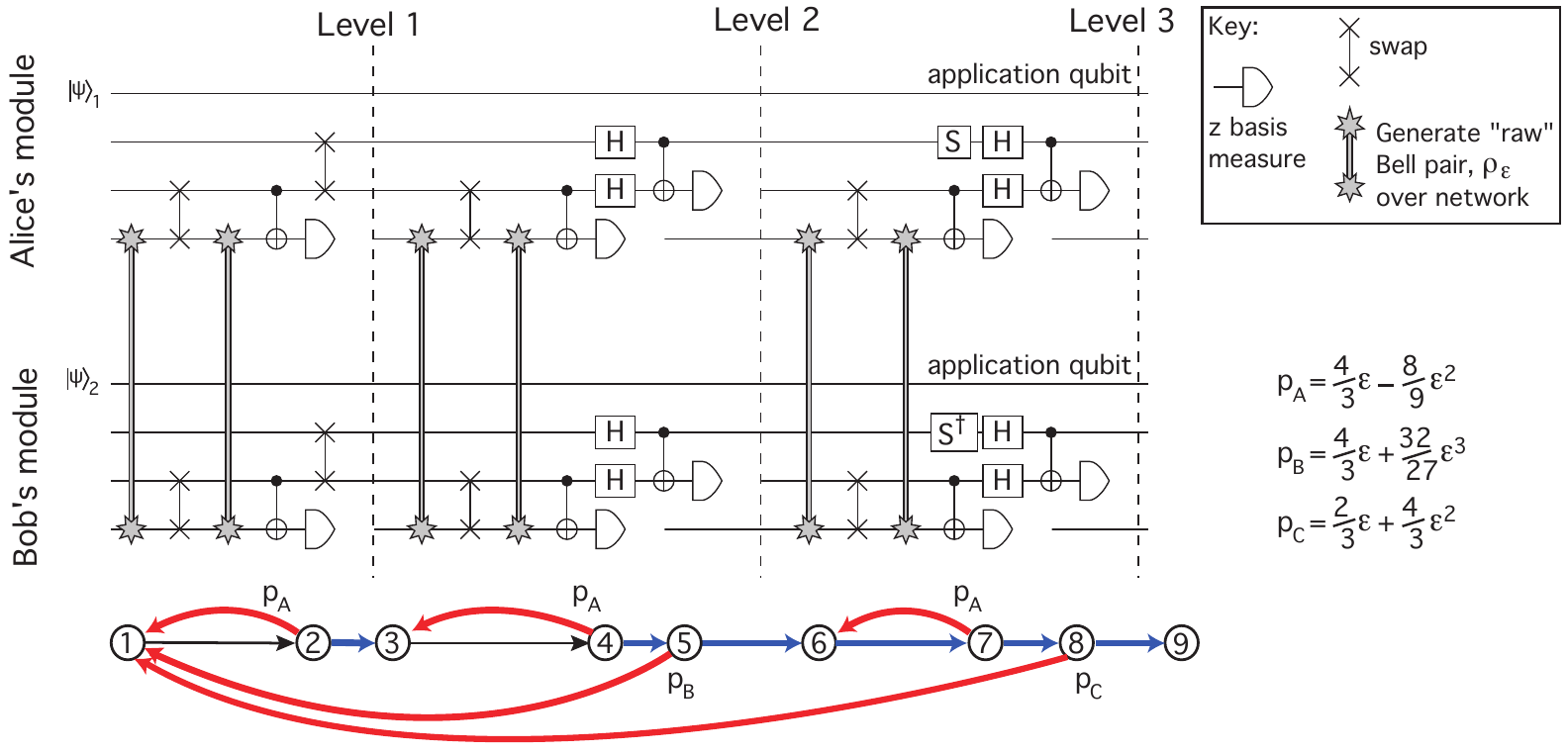}

\caption{(Top) Multi-level purification circuit using generation of ``raw'' Bell pairs  $\rho_{\epsilon}\sim(\frac{\epsilon}{3},\frac{\epsilon}{3},\frac{\epsilon}{3})$, Clifford gates and measurements as primitive operations. For clarity, the application qubits $\ket{\psi}_A$ and $\ket{\psi}_B$ are also shown even though they do not take part in purification. The purification can be terminated at different stages of the circuits, which define the purification level. Level 1, Level 2 and Level 3 protocols produce purified Bell states with the infidelities of $\sim\frac{2}{3}\epsilon$, $\sim\frac{8}{9}\epsilon^2$ and $\sim \frac{2}{9}\epsilon^2$ respectively. (Bottom) The Markov chain showing the progression along the purification protocol; $p_A$, $p_B$ and $p_C$ denote the probabilities of odd parity measurement result at various stages of the protocol. The dependence of these probabilities on the network noise $\epsilon$ can be obtained by direct computation and the first two terms in their Taylor expansions are indicated on the diagram. 
\label{fig:dist6}}
\end{figure*}

We assume that the raw entangled state is of the form of a depolarized
Werner state

\begin{equation}
\rho_{\epsilon}=\left(1-\epsilon\right)\Phi^{+}+\frac{\epsilon}{3}\Phi^{-}+\frac{\epsilon}{3}\Psi^{+}+\frac{\epsilon}{3}\Psi^{-}, \label{eq:werner}
\end{equation}
where $\epsilon\in[0,0.5)$ and states $\Phi^{+}$, $\Phi^{-}$, $\Psi^{+}$
and $\Psi^{-}$ are the standard Bell states. Here we have chosen
$\Phi^{+}$ as the desired Bell state (obviously, it requires only a single-qubit rotation to transform any Bell state to another, so we are not limiting ourselves to any particular entanglement generation protocol by assuming $\Phi^+$). The fidelity of $\rho_{\epsilon}$
is $\mathrm{Tr}\sqrt{\sqrt{\rho_{\epsilon}}\Phi^{+}\sqrt{\rho}_{\epsilon}}=1-\epsilon$.  A state given by Eq. (\ref{eq:werner}) is fully depolarized i.e. its errors are evenly distributed across the $X$, $Y$ and $Z$ error channels. As noted above, in the context of purification, a fully depolarized state input state is a conservative assumption since structured noise can be beneficially exploited in purification. 

The most basic purification circuit takes two pairs $\rho_{\epsilon}$
and produces a state $\tilde{\rho}_{\epsilon}$ of fidelity greater
than $1-\epsilon$. It consists of two cNOT gates and a parity measurement. This purification protocol is the first part (Level 1) of the circuit shown in Figure \ref{fig:dist6}. We will denote
the map describing this purification protocol by $F:\textbf{Q\ensuremath{\otimes\textbf{Q}\rightarrow\textbf{Q}}}$,
where $\textbf{Q}$ is a set of two qubit density matrices. Measurements
are in the standard basis and are postselected to have the same parity.
The resulting state $\tilde{\rho}_{\epsilon}^{(1)}$ is given by 

\begin{widetext}
\begin{equation}
\tilde{\rho}_{\epsilon}^{(1)}\equiv F[\rho_{\epsilon},\rho_{\epsilon}]=\left(1-\frac{2}{3}\epsilon-\frac{2}{3}\epsilon^{2}\right)\Phi^{+}+\left(\frac{2}{3}\epsilon+\frac{2}{9}\epsilon^{2}\right)\Phi^{-}+\frac{2}{9}\epsilon^{2}\Psi^{+}+\frac{2}{9}\epsilon^{2}\Psi^{-}+O(\epsilon^{3}).\label{eq:phitild}
\end{equation}
\end{widetext}
The fidelity of $\tilde{\rho}_{\epsilon}^{(1)}$ is $1-\frac{2}{3}\epsilon+O(\epsilon^{2})$. To achieve further improvements in fidelity, one 
constructs a tiered purification protocol, where a level consists
of a single application of map $F$ and the outputs from a given level may
be inputs at a higher level. As we will see shortly, between each
stage of the process it may be necessary to perform local rotations. 

We now systematically construct such multilevel purification
protocols. First, we introduce a shorthand notations where the state

\begin{equation}
\rho=(1-r_{1}-r_{2}-r_{3})\Phi^{+}+r_{1}\Phi^{-}+r_{2}\Psi^{+}+r_{3}\Psi^{-},\label{eq:gena}
\end{equation}
is represented by a tuple 

\begin{equation}
\rho\sim(r_{1},r_{2},r_{3}).\label{genb}
\end{equation}
By direct computation one finds that (to lowest order) the effect
of a single iteration of map $F$ on states of general form (\ref{genb})
is

\begin{equation}
F\left[(r_{1},r_{2},r_{3}),(s_{1},s_{2},s_{3})\right]\sim(r_{1}+s_{1},r_{2}s_{2}+r_{3}s_{3},r_{2}s_{3}+r_{3}s_{2}).\label{eq:Fmap}
\end{equation}

Thus, for instance, applying equation (\ref{eq:Fmap}) on a pair of
states $\rho_{\epsilon}\sim\left(\frac{\epsilon}{3},\frac{\epsilon}{3},\frac{\epsilon}{3}\right)$,
produces a state $\tilde{\rho}_{\epsilon}^{(1)}\sim\left(\frac{2}{3}\epsilon,\frac{2}{9}\epsilon^{2},\frac{2}{9}\epsilon^{2}\right)$
in agreement with equation (\ref{eq:phitild}). From equation (\ref{eq:Fmap})
we can see that the map $F$ suppresses the contributions from the
$\Psi^{\pm}$ modes but increases the contribution of the $\Phi^{-}$
channel. Further iterations of the map $F$ will decrease the fidelity
of the resulting state due to the concentration of noise in the $\Phi^{-}$
mode. In order to continue to improve the fidelity with successive
applications of $F$ one can permute the $\Phi^{-}$, $\Psi^{+}$
and $\Psi^{-}$ modes by applying local rotations. The three modes
$\Phi^{-}$, $\Psi^{+}$ and $\Psi^{-}$ can be described in terms
of single-qubit Pauli errors on the noise-free mode $\Phi^{+}$

\begin{eqnarray*}
\Phi^{-} & = & (I\otimes Z)\Phi^{+}(I\otimes Z)\\
\Psi^{+} & = & (I\otimes X)\Phi^{+}(I\otimes X)\\
\Psi^{-} & = & (I\otimes Y)\Phi^{+}(I\otimes Y).
\end{eqnarray*}

These modes can be permuted using any single-qubit Clifford group
operations which leave $\Phi^{+}$ invariant. These operations form
a representation of the dihedral group $D_{3}$, and in particular
form a group of order 6; they are generated by the operation $g_{1}\equiv H\otimes H$
and the operation $g_{2}=S^{\dagger}\otimes S$, where $S=\exp\left(-i\frac{\pi}{4}Z\right)\propto\mathrm{diag}(1,i)$.

With this in mind, we can see that one can form a higher fidelity
state $\tilde{\rho}^{(2)}$ by applying $F$ on two states $\tilde{\rho}^{(1)}=F[\rho_{\epsilon},\rho_{\epsilon}]$
if they are first locally rotated using $g_{1}$.

\begin{align}
\tilde{\rho}^{(2)}&= F[g_{1}\tilde{\rho}^{(1)}g_{1}^\dagger,g_{1}\tilde{\rho}^{(1)}g_{1}^\dagger] \nonumber \\ 
 \sim & F\left[\left(\frac{2}{9}\epsilon^{2},\frac{2}{3}\epsilon,\frac{2}{9}\epsilon^{2}\right),\left(\frac{2}{9}\epsilon^{2},\frac{2}{3}\epsilon,\frac{2}{9}\epsilon^{2}\right)\right] \nonumber \\ 
 \sim & \left(\frac{4}{9}\epsilon^{2},\frac{4}{9}\epsilon^{2},\frac{8}{27}\epsilon^{3}\right).
\end{align}
The fidelity of this state is $1-\frac{8}{9}\epsilon^{2}+O(\epsilon^{3}).$
This purification map that produces state $\tilde{\rho}^{(2)}$ is represented by the Level 2 of the circuit shown in Figure  \ref{fig:dist6}. 

Finally, we construct a third level to our purification protocol that uses $\tilde{\rho}^{(2)}$ and $\tilde{\rho}^{(1)}$, with suitable local rotations, to produce a purified
state $\tilde{\rho}^{(3)}$. The Level 3 purification protocol is represented by the whole circuit shown in Figure \ref{fig:dist6}. To the lowest orders in $\epsilon$ the state
$\tilde{\rho}^{(3)}$ is given by

\begin{align}
\tilde{\rho}^{(3)} & =  F\left[g_{1}g_{2}\tilde{\rho}^{(2)}g_{2}^\dagger g_{1}^\dagger,g_{1}\tilde{\rho}^{(1)}g_{1}^\dagger\right]\nonumber\\
 & \sim F\left[g_{1}g_{2}\left(\frac{4}{9}\epsilon^{2},\frac{4}{9}\epsilon^{2},\frac{8}{27}\epsilon^{3}\right)g_{2}^\dagger g_{1}^\dagger,g_{1}\left(\frac{2}{3}\epsilon,\frac{2}{9}\epsilon^{2},\frac{2}{9}\epsilon^{2}\right)g_{1}^\dagger\right]\nonumber\\
 & \sim F\left[g_{1}\left(\frac{4}{9}\epsilon^{2},\frac{8}{27}\epsilon^{3},\frac{4}{9}\epsilon^{2}\right)g_{1}^\dagger,\left(\frac{2}{9}\epsilon^{2},\frac{2}{3}\epsilon,\frac{2}{9}\epsilon^{2}\right)\right]\nonumber\\
 & \sim F\left[\left(\frac{8}{27}\epsilon^{3},\frac{4}{9}\epsilon^{2},\frac{4}{9}\epsilon^{2}\right),\left(\frac{2}{9}\epsilon^{2},\frac{2}{3}\epsilon,\frac{2}{9}\epsilon^{2}\right)\right]\nonumber\\
 & \sim \left(\frac{2}{9}\epsilon^{2},\frac{8}{27}\epsilon^{3},\frac{8}{27}\epsilon^{3}\right). \label{level3channels}
\end{align}

The complete Level 3 purification circuit uses three pairs
of raw states $\rho_{\epsilon}$ of infidelity $\epsilon$ to produce
a state $\tilde{\rho}^{(3)}$ of infidelity $\sim\frac{2}{9}\epsilon^{2}$.
If we were to add to the resources another pair of raw states $\rho_{\epsilon}$
then it is possible to produce a state $\tilde{\rho}^{(4)}$ of infidelity
$\sim\frac{16}{27}\epsilon^{3}$. However, in the current paper, we will not
analyse maps that produce states of infidelity of order $\epsilon^3$ and higher. We presently find that states produced by the Level 3 purification protocol are already of fidelity that is high enough for quantum communication and computing applications. 

Note that each additional stage of purification requires generation of an additional two raw Bell pairs, while introducing a new {\it tier} to the process would double the total requirements. Since with current technology photonic entanglement is a slow operation, protocols significantly more complex that that described above would likely lead to unacceptable slowdown to the rate at which remote gates can be applied between the application qubits in separate modules.

\begin{figure*}[t]
\centering
\includegraphics[scale=0.43]{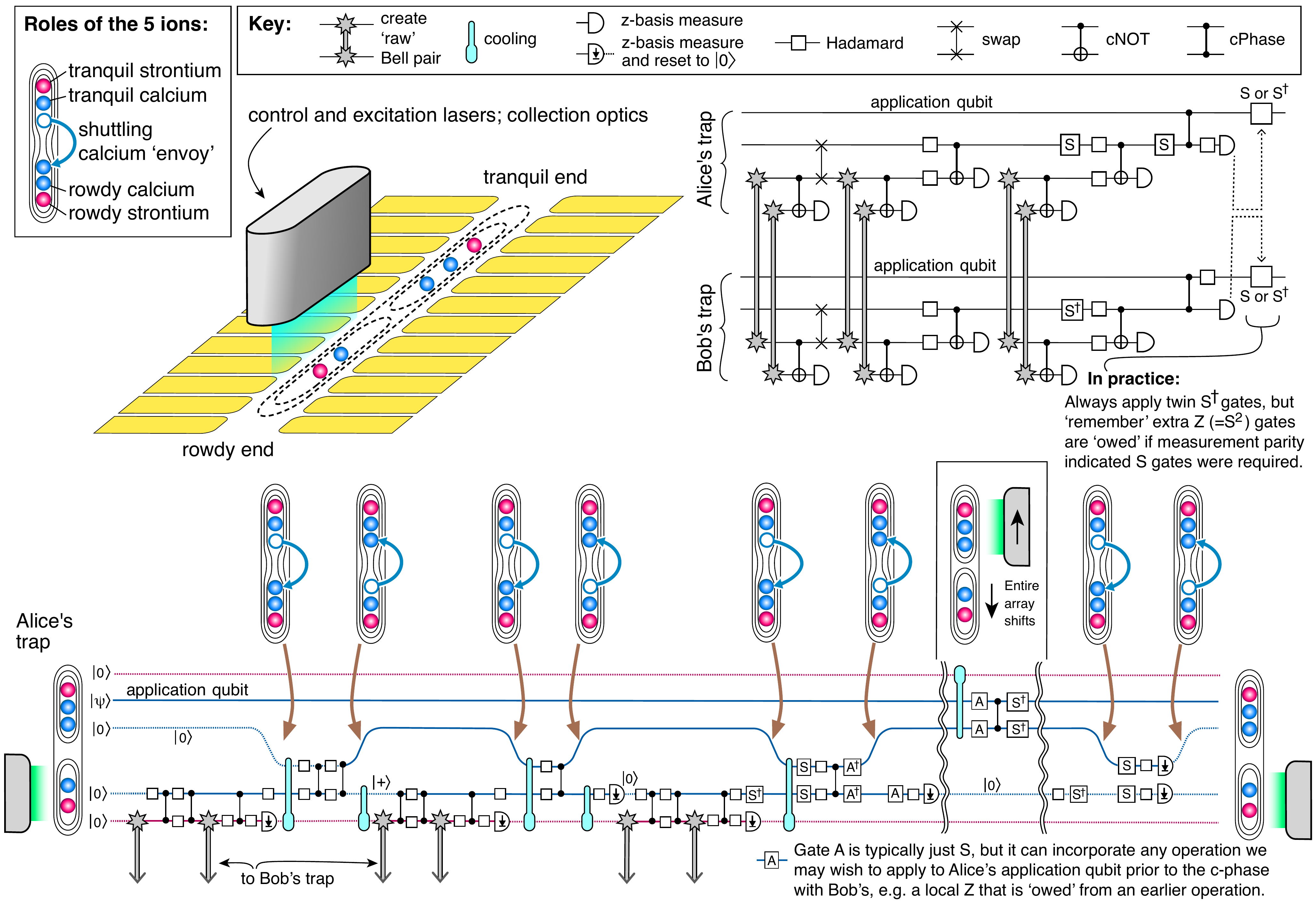}

\caption{Diagram showing the physical layout and the operation of the proposed ion trap quantum network node. The device holds two mixed species ion crystals in separate potential wells. The Sr$^+$ ions is used to generate ion/photon entanglement and for cooling. The Ca$^+$ ions are used in the purification protocols, for storing quantum information (application qubit) as well as mediating the gates between the application qubits of photonically linked nodes. All of the control, measurement and excitation optics is concentrated in one of the segments, which reduces the noise acting on the spectator qubits but introduces the need for ion shuttling.   
\label{fig:ionmotion}}
\end{figure*}

 The purification protocol is postselective - if a measurement produces an odd parity result then the protocol has to be repeated from a particular point. Thus in practice the number of raw entanglement pairs needed to complete a protocol is not fixed. The Markov chain representing the progression along the purification protocol is shown below the circuit diagram in Figure 2. 

A few remarks about the optimality of the above purification circuits are in order. 
Our choice of the primitive 2-to-1 purification protocol is strongly
motivated by the setting in which it is to be applied. As we prefer to
perform nearest-neighbour operations (which themselves
are noisy) within a linear array, it is important to minimise the
number of operations to be performed to reduce the propagation of
errors. Furthermore, as we consider traps with very few ions, it is
important to limit the number of qubits to be stored simultaneously. Given these considerations, the
simplicity of the 2-to-1 purification is advantageous.
We then require only that the raw entangled states have little enough
noise that each round of purification may succeed with high
probability.

It remains to consider whether we may obtain further improvements in
the noise reduction, given the same device complexity but using maps other than the particular
arrangements of Hadamard gates and $\pi$/4 gates which we describe. Our choices of transformations are optimal over the
set of Clifford gates: operations from the Clifford group only serve
to permute the Pauli noise channels, and our operations are chosen to
optimise the rate at which these noise channels are suppressed upon
success. Any choice of non-Clifford gates will at best mix the Pauli
noise channels prior to purification, and at worst introduce more
noise channels which are not described by Pauli operators, reducing
the (admittedly small) probability of cancelling the noise from
different noisy entangled states, without increasing the probability
of success in purification. This leads us to conclude that the approach we have taken
is the best choice for purification in our setting.

\section{Physical layout of the device \label{sec:layout}}

We now combine the designed purification circuit shown in Figure 2 with the general ion trap consideration of Section II to produce a detailed blueprint for the ion trap quantum network node and its operational steps. This is presented in Figure~\ref{fig:ionmotion}.

\begin{figure*}
\includegraphics{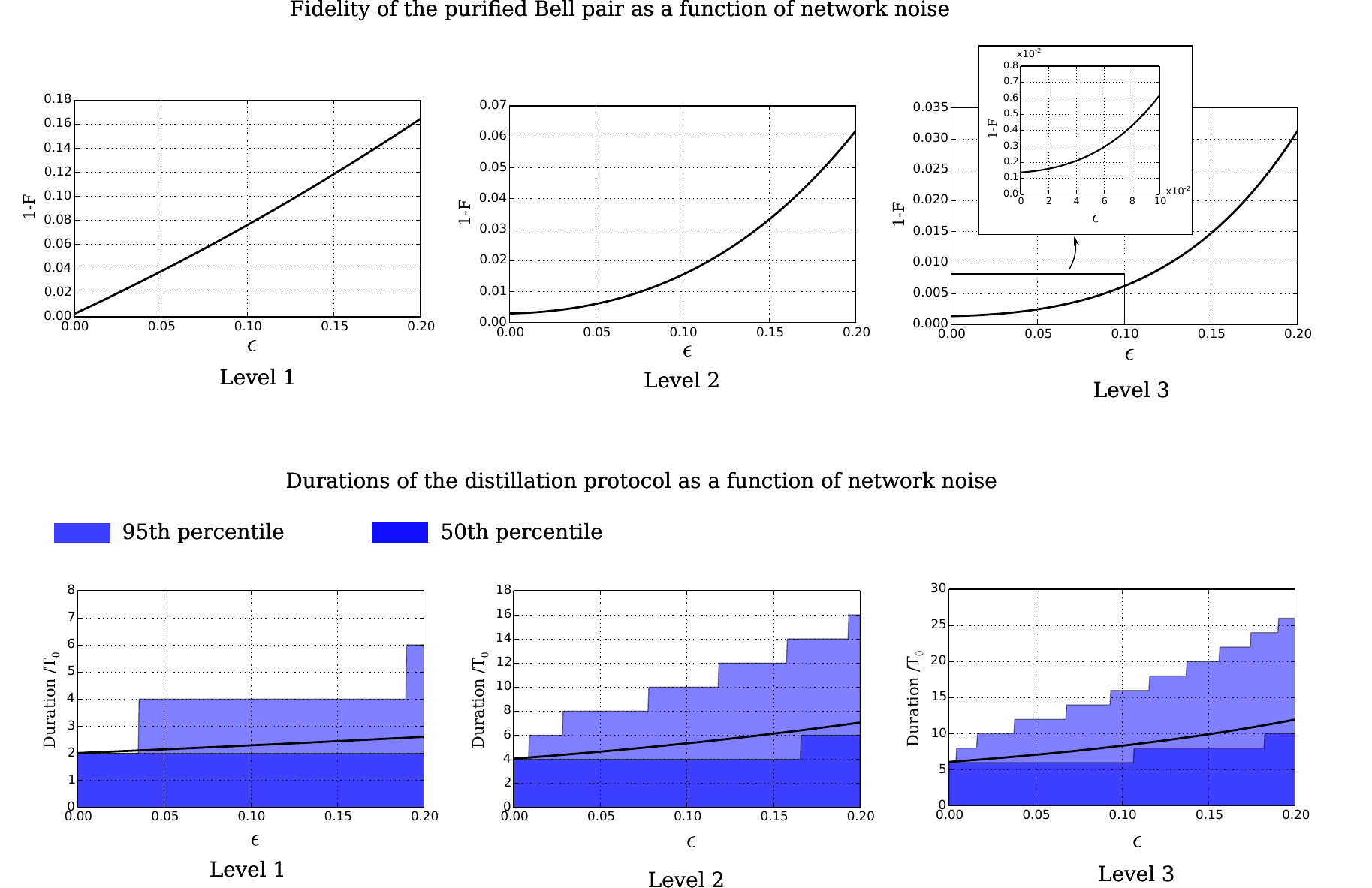}
\caption{(Top) The infidelity $1-\mathcal{F}$ of a state produced by Level 1, Level 2 and Level 3 purification circuits as a function of the infidelity of the input raw entangled state $\epsilon$. The simulated purification circuit is given by Figure \ref{fig:ionmotion} and the assumed error rates for the single qubit rotation, two qubit rotations and measurements are respectively $p_{1}=1\times10^{-6}$, $p_{2}=0.001$ and $p_{m}=0.0005$.
(Bottom) The probabilistic duration of the Level 1, Level 2 and Level 3 purification protocols as a function of the infidelity of the input raw entangled state $\epsilon$. The duration is given in terms of the time it takes to generate a single raw entangled pair, $T_0$, and the all other operations, such as qubit rotations and ion shuttling, are assumed to be instantaneous. \label{fig:fidelity}}
\end{figure*}

The allowed primitive operations in the device are ion shuttling operations (splitting, joining and moving ion crystals in a linear array), raw photonic entanglement generation, local qubit rotations, the symmetric two-qubit phase gate cPhase, measurements, and crystal cooling. (Note that the basic two-qubit operation may not in practice be the cPhase, but rather e.g. a M\o lmer-S\o rensen gate; however by suitably replacing the adjacent pairs of Hadamard gates by another symmetric pair of single-qubit gates, they are made equivalent).  Figure~\ref{fig:ionmotion} shows the suggested device layout with the explicit purification and application sequence. At any time the module contains two mixed-species  ion crystals; there will be two ions in one potential well, and three in the other. In this paper we assume the species are $^{43}$Ca$^+$ and Sr$^+$, however there are are of course other suitable possibilities.  The two Sr$^+$ ions are used for cooling and/or photonic entanglement generation. The three $^{43}$Ca$^+$ ions are used for purification, storing quantum information and mediating the gates between the application qubits of separate modules. The control, excitation and collection optics are all focused on one trap region only, i.e. it targets only one of the ion crystals. Typically the targeted region is the rowdy end; when the time comes for the envoy qubit to entangle with the application qubit, the entire trap potential shifts (without any change to the relative positions of the ions) so that the laser control systems now target the tranquil end; once the gates there are implemented, the potentials shift back. 

Note that in the fully compiled circuit it is never necessary to differentially target one ion over another of the same species in the same zone; in fact zones are under {\em global control} in the sense that control pulses target entire zones and, where there is more than one ion of the relevant species, both respond: we therefore restrict ourselves to symmetric gates cPhase gate and $G\otimes G$, where $G$ is any single qubit rotation. 
This negates the issue of cross talk {\it within} a given zone, leaving us only concerned with the possibility of accidental excitation (by scattered laser light or emitted photons) of ions in the other zone; given that the zone separation could be of the order of a centimetre if need be, this source of error should be easily made negligible. 

\section{Performance of two connected modules \label{sec:twoConnectedMods}}

In this Section, we numerically evaluate the performance of the designed purification protocols assuming realistic level of gate noise. Single qubit
noise is modelled by a perfect gate followed by a trace preserving
noise process

\begin{equation}
\mathcal{N}_{1}(\rho)=(1-p_{1})\rho+\frac{p_{1}}{3}\left(X\rho X+Y\rho Y+Z\rho Z\right),
\end{equation}
where $X$, $Y$ and $Z$ denote the Pauli matrices.

Two qubit noise is modeled by perfect gate followed by a noise process

\begin{equation}
\mathcal{N}_{2}(\rho)=(1-p_{2})\rho+\frac{p_{2}}{15}\sum_{A,B}\left(A\otimes B\right)\rho\left(A\otimes B\right)^{\dagger},
\end{equation}
where operator $A\in\{I,X,Y,Z\}$ acts on the first qubit, and similarly
$B$ acts on the second qubit but the case $I\otimes I$ is excluded
from the sum. Note that a given experimental system will have noise that deviates from an even distribution of errors over all channels (see e.g. Ref.~\cite{Ballance2015a}), but by making this assumption we ensure that all error types are corrected.

Given the measurement error rate $p_m$,
a particular outcome $q \in \{0,1\}$ corresponds to the intended projection $P_q$ applied
to the state with probability $(1-p_m)$ and the opposite projection $P_{\bar{q}}$ applied with probability
$p_m$. The superoperator describing the measurement is thus

\begin{equation}
\mathcal{P}_{m}(\rho)=(1-p_m)\ket{q}\bra{q}+p_m\ket{\bar{q}}\bra{\bar{q}}.
\end{equation}

The fidelity of the purified state as a function of $\epsilon$ is
shown in Figure \ref{fig:fidelity} for Level 1, Level 2 and Level 3 purification protocols.
The values chosen for the intra-module error rates correspond to the values reported in recent ion trap experiments \cite{Ballance2015a,Harty2014}, $p_{1}=1\times10^{-6}$, $p_{2}=0.001$ and $p_{m}=0.0005$. Note the quoted single qubit fidelity of $10^{-6}$ is from a microwave controlled operation, and such gates have yet to be fully localised even to the scale of one of our trap regions (rather than encompassing both). However, laser controlled single-qubit gates are also very high fidelity, $4\times 10^{-5}$ in Ref.~\cite{Wineland2016}; adopting such a number instead would not alter any of our results appreciably. Figure \ref{fig:fidelity} also displays the probabilistic running times of the purification protocols calculated by numerically simulating the Markov chain in Fig. \ref{fig:dist6}.

\begin{figure*}[t]
\centering
\includegraphics[scale=0.78]{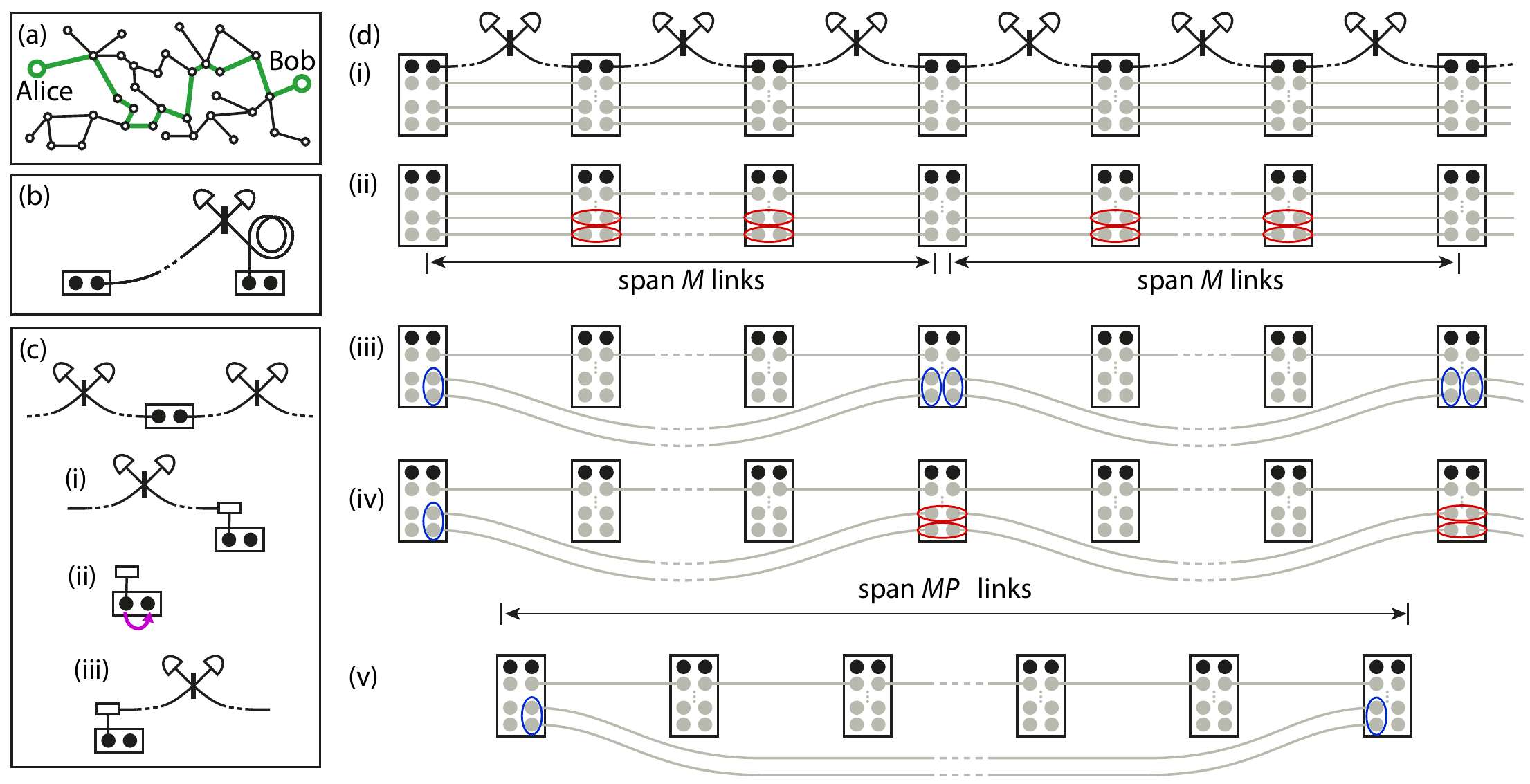}
\caption{Communication over an arbitrary network of modules (a). Alice and Bob need not trust the other network nodes. (b) Entanglement-inducing measurements could be co-located with the modules. (c) Modules can be as simple as the five ion device, although then entanglement with left and right neighbours must be created in separate steps. (d) A more ideal device would have additional internal memory, allowing purification (i), (iii) and (v) to alternate with fusion (ii) and (iv) in a near deterministic fashion. As described in the main text, a path length of $2,400$~km may be reached with two tiers of purification.
\label{fig:commsNet}}
\end{figure*}

\section{Application: Communication \label{sec:communication}}

In this section we consider how a purification module as described above can be used to distribute high quality Bell pairs between remote locations in a network. Performance will be estimated in the context of quantum key distribution. However the module is not limited to that application; being capable of purifying, storing and processing entangled states, it is a general enabler for communications applications and can be thought of as a universal communication node.  

We will consider a relatively naive use of the purification module to act as a repeater, so that a chain of such modules spans the distance between two remote parties. This will be a `first generation' approach in the sense of Ref.~\cite{LiangmultiGenerationsOfRepeater}. We estimate some performance characteristics in this simple scenario, and indicate where a more sophisticated approach based on code states may become preferable. 

In our scenario Alice and Bob are at two remote points in a quantum network, and they wish to use the network to generate Bell states so as to create a shared secret key known only to themselves. Being part of a network they do not have a direct connection between them, but they can identify a path (or paths) involving a number of intermediate nodes. For simplicity we will assume that they identify a single path and make exclusive use of the nodes along it until they have succeeded in their task, see Fig.~\ref{fig:commsNet}(a). Obviously, generalisations are possible involving using multiple paths and/or sharing node functionality with other network users. Alice and Bob will use the nodes along the path in order to generate Bell pairs that they alone share; they need not trust the operators of those intervening nodes.

Suppose that each link along the path is an ion trap device as outlined previously. We assume that a frequency conversion technology is used in order to translate single photons from the ion's native emission frequency to a telecoms frequency $~1550$~nm, so that transmission over distance through fibre is feasible. We also assume that the ion trap modules alternate with measurement modules along the chain as in Fig.~\ref{fig:commsNet}(d) (although it is possible that a measurement system can be co-located with each ion trap Fig.~\ref{fig:commsNet}(b)).

Each module could be as simple as the five-ion device detailed above, in which case entanglement would be created and purified first with the neighbour to one side, then stored in the `application qubit' (which acts simply as a memory) while entanglement is  created and purified with the neighbour to the other side (Fig.~\ref{fig:commsNet}(c)). This could suffice for a smaller scale network. However in Figure~\ref{fig:commsNet}(d) a series of grey qubits are indicated; these are additional memory ions which are not essential but have the effect of increasing efficiency: they store purified Bell pairs so as to provide a `buffer' between Bell creation and consumption. Provided that the rate of consuming the Bell pairs is set to be somewhat slower than the average creation rate, the buffers will tend to replenish and the system can operate in a near deterministic way, rather than having the entire device wait for the slowest link. The result will be an increase in bit rate by a factor that is logarithmic in the chain length; this is likely to be a worthwhile enhancement and will require a third zone in addition to the two indicated in Fig.~\ref{fig:ionmotion}. Since this zone is purely for storage, it should not add greatly to the complexity. 

All nodes are continuously seeking to create purified Bell pairs with their immediate neighbours, Fig.~\ref{fig:commsNet}(d)(i). Periodically, certain modules will act to fuse a Bell qubit that is shared with a left neighbour with a Bell qubit shared to the right. This is achieved by performing a Bell-basis measurement on the local pair, recording the measurement outcome and communicating it classically to one of the neighbouring modules; that module can apply a local single-qubit rotation to complete the process. (Note that a complete Bell-basis measurement is easily achieved using, for example, a control-NOT between the two qubits, followed by a Hadamard on the controlling qubit and then measurement of each qubit separately in the $z$-basis.) As shown in Fig.~\ref{fig:commsNet}(d)(ii) these fusion operations can happen in $M-1$ consecutive modules so as to fuse together $M$ Bell pairs in a single step. The result is a long-range Bell pair shared between the end-point nodes. Any noise present in the original Bell pairs will contribute to the noise in the newly fused, long-range pair. Therefore these long-range pairs are purified, before the same process is repeated to join fuse them into very long-range pairs.

The costs of this process can be estimated as follows: Suppose that the infidelity in the raw entanglement process is $10\%$ (and this must include the effects of the frequency conversion technology). Take the initial purification to be a Level 3 process according to Fig.~\ref{fig:dist6}. This will produce Bell pairs will an infidelity of $0.6\%$, and moreover the noise will be largely in a single channel (see Eqn.~(\ref{level3channels}) and Table~\ref{table:longRange}). The average time cost will correspond to the creation of $8.34$ `raw' Bell pairs. When a chain of $M$ such Bell pairs are now fused together by high grade local operations (Fig.~\ref{fig:commsNet}(d)(ii)), the result will be a Bell pair whose infidelity is greater by a factor of approximately $M$. For $M=12$, numerical simulation produces the numbers shown in Table~\ref{table:longRange}. Now suppose that a Level 2 purification is performed on these pairs. Because of the structure in the noise, level two suffices to reduce the infidelity back below $0.6\%$ as shown in the Table. This necessarily consumes at least $4$ of the long range pairs, and because of the failure possibility the average cost is in fact $4.77$. Finally the long range pairs are again combined, with $P$ of them being fused into very-long-range pairs. If $P=12$, then we have a total range of $12\times 12=144$ modules. Finally performing another Level 2 purification yields final Bell pairs with infidelity $0.55\%$, at an average cost $4.78$ input pairs. We see that we can characterise this process by saying that we suffer a reduction in the rate of pair distribution by a factor of $\sim 4.8$ every time we increase the range by a factor of $12$. The process could be continued to reach longer chains, using the same rule.

\begin{table}[ht]
\centering 
\begin{tabular}{| c | l | l |} 
\hline
\ stage\ & $\ \Phi^+$ (fidelity)\ & \ error channels (decreasing order)\ \\ [0.5ex] 
\hline 
i &\  0.993817\ 	&\ 0.00443,\  0.000957,\  0.000796 \\ 
ii &\ 0.922 	&\ 0.052,\  0.01396,\  0.0123\\
iii &\ 0.994154 	&\  0.00433,\ 0.00103,\  0.000487  \\
iv &\ 0.925 	&\ 0.0511,\ 0.0146,\  0.00888   \\
v &\ 0.99450 	&\ 0.0044,\ 0.000681,\  0.000422  \\ [1ex] 
\hline 
\end{tabular}
\caption{Creating long range entanglement. Note that the initial (short range) purified fidelity at stage (i) is recovered at stages (iii) and (v). These data are obtained using a  local two-qubit gate fidelity of $99.9\%$, but neglecting single qubit errors (which have been realised at far higher fidelity). During purification processes (iii) and (v), local operations are used to permute the three erroneous Bell states so that the state which will escape the next purification has the lowest probability. } 
\label{table:longRange} 
\end{table}

Suppose that the modules are spaced apart by $d$ kilometres, and that  standard silica telecoms fibre (Corning SMF-28) is used with a photon loss rate of order $0.17$~dB per kilometre. Then the loss in reaching the measuring station mid-distance between modules, Fig.~\ref{fig:commsNet}(d)(i) will be $(0.085)d$~dB. However, it is likely that a procedure involving detecting two photons would be employed in order to alleviate the demands for interferometric stability. Therefore the photon loss probability will impact the success rate quadratically, so that it falls as $(0.17)d$~dB. We might reasonably assume that $d$ is chosen so that the total success rate only falls by a factor of $2$ (on top of all other loss mechanisms including collection and detector inefficiency, losses in links and within the frequency conversion system). Then $d=17$~km is an appropriate spacing, and our chain of $12\times 12=144$ modules spans more than $2,400$~km. But to reach this range we have used several tiers of purification: a Level 3 purification at the initial stage of entanglement generation between modules, costing $8.34$ raw pairs per success, and then two additional purification phases (Fig.~\ref{fig:commsNet}(c)(iii)  and (c)(v)) each of level 2, taking $4.77$ and $4.78$ input pairs, respectively, to produce one upgraded pair. The total factor between the raw Bell pairs and the purified remote pairs is therefore $190$. (Note that a further factor of $~4.8$ would allow us another factor of twelve in separation, reaching $29,000$~km and so exceeding the distance between any two points on Earth's surface).

To the factor of $190$ we must add the cost of translating the remote Bell pairs into shared secret bits, i.e. the QKD protocol itself. The Bell pairs we have created have higher fidelity than those considered in the relevant literature, and moreover our noise is concentrated into specific channels. Therefore while it is not possible to identify an exact cost for this stage without further study, it is reasonable to hope that this cost can be somewhat less that those that have been computed to date: Inspecting Table 1 in Ref.~\cite{Tomamichel2012} one might expect the ratio between rates of shared key creation and Bell pair distribution to be between 0.2 and 0.6, depending on how large a key is generated (larger keys being more efficient to generate). 

Ultimately, then, we may see nearly three orders of magnitude reduction between the rate of `raw' entanglement generation between neighbouring modules, and the rate of generating secure bits between users $2,400$~km apart. Given that key lengths of $10^4$ bits may be necessary for practical QKD, one would require a raw entanglement generation rate of $100$~kHz to create such a key in $100$~seconds. This is very demanding given that raw entanglement rates {\it in the lab} are presently four orders of magnitude slower. However, it is reasonable to suppose that multiple ion trap devices can and would be implemented within each repeater station; since each would be independent from the others, this would not increase the technical sophistication and indeed the expensive components such as control lasers could be used as a common resource for multiple traps. Through such an approach, together with anticipated improvements in the efficiency of collecting light from ions, it may be possible to reach the communication rates described here. 

Note that once one considers having multiple ion trap devices within each repeater station, the possibility arises that one could interlink those devices locally; then it would become possible to use an encoding such as Raussendorf's 3D cluster state~\cite{Raussendorf20062242} to fault tolerantly transmit entangled pairs~\cite{fowlerSurfaceCodeComms2010}, with each repeater forming a `sheet' of the structure. One could employ a variant of the approach described in Ref.~\cite{YingCommunicate} to achieve transmission rates that do not deteriorate with distance (at the cost of considerably greater complexity in the repeaters). Moreover such a system, being a `third generation' repeater~\cite{LiangmultiGenerationsOfRepeater}, need not be limited by classical communication times -- we conclude by assessing the significance of this limit for our naive approach.

The finite nature of the speed of light leads to a bound on the entanglement rate: In order to know whether an entanglement attempt has succeeded or failed one must wait for light to travel from the ion to the detector system, and the return of a classical signal. If the attempt has failed, the ion cannot be reset for another attempt until this information is received; this therefore puts a limit on the attempt rate of $c/d$ where c is the speed of light. In dense urban networks this might not be an issue, since $d$ may be less than a kilometre. But for our long range network where we have assumed $d=17$~km, the implied maximum cycle rate is $18$~kHz, a factor of $26$ below the $470$~kHz rates that have been used in the lab in entanglements~\cite{Hucul2015}. The obvious solution is to reduce the separation $d$ between repeater stations, but this is expensive and will imply that more purification is needed for a given total distance. A more advanced solution would be to have multiple ions available for `raw' entanglement, such that each acts briefly as memory while the results of its latest entanglement attempt are awaited. Within this time several other ions would begin the process of entanglement generation, sequentially. 

Thus for long range networks an ideal device might employ the same central two zones are as described in Section~\ref{sec:layout}, but peripherally there would be one zone for passive memory ions (the grey circles in Fig.~\ref{level3channels}) and another region containing several ions that are dedicated to obtaining raw entanglement.

\color{black}

\section{Application: Fault tolerant computing \label{sec:computing}}

\begin{figure}[t]
\centering
\includegraphics[scale=0.7]{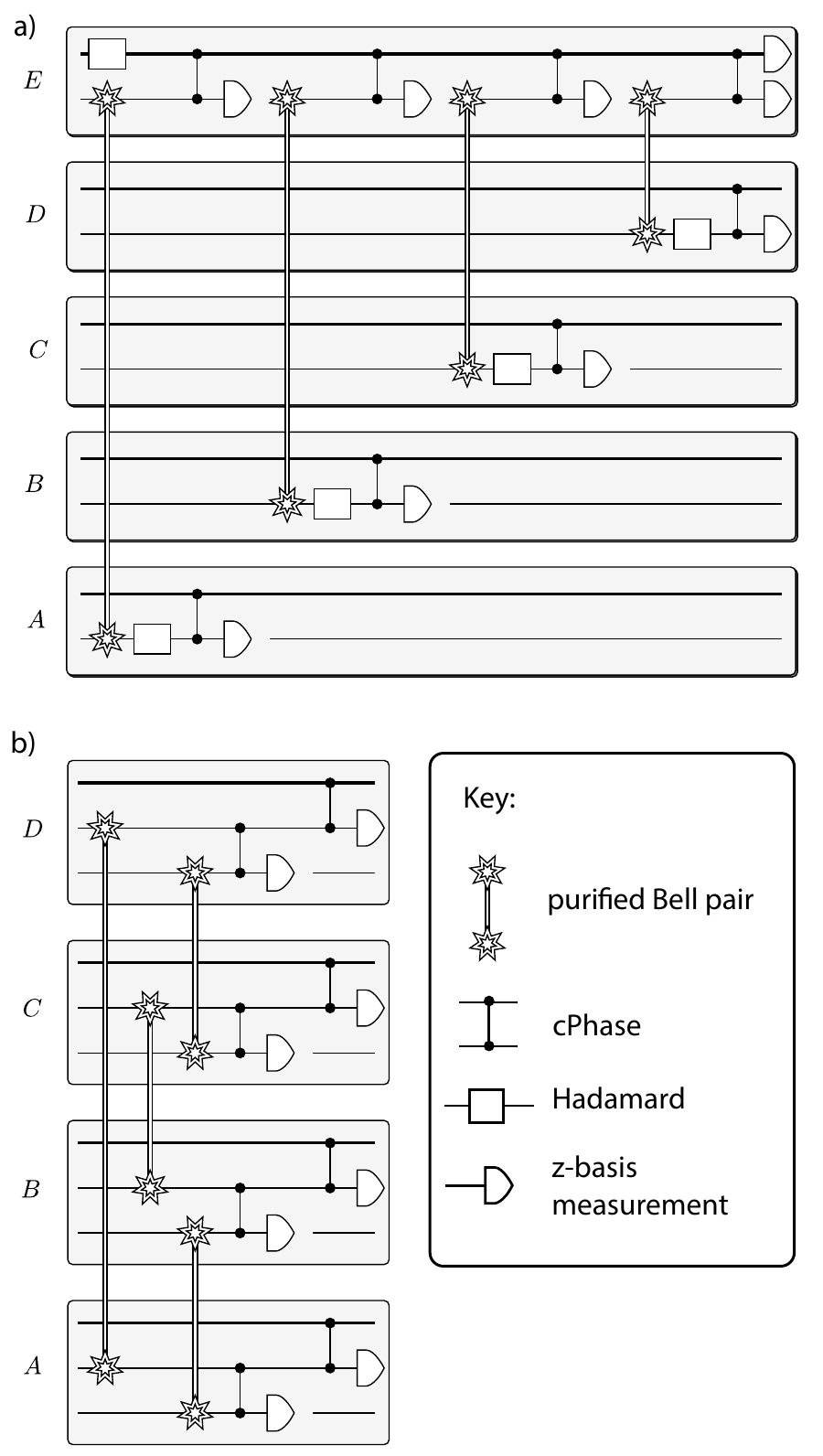}
\caption{(a) Circuit implementing a parity measurement between application qubits in
nodes $A$, $B$, $C$ and $D$ using an ancilla node $E$ and two-qubit gates enabled by purified Bell pairs. (b) Circuit implementing parity measurements between nodes  $A$, $B$, $C$ and $D$ using a shared four qubit GHZ state. \label{fig:ancilla_stabilizer_circuit}}
\end{figure}

We now consider the use of our modules in the context of scalable, fault tolerant quantum computing. We will evaluate the performance of our five-ion module as described earlier, as a building block of an  architecture that uses the toric quantum error correcting code. The application qubit within each module now represents one ``data qubit'' of the toric code. We numerically simulate the code and determine the error correction thresholds in terms of the network noise. 

The toric code involves repeatedly measuring {\it stabilisers}, which correspond in practice to parity measurements on groups of four data qubits.
The basic repeating cycle of the computer involves alternating these measurements with
Hadamard rotations to switch between the $x$ and $z$ basis. In addition to preserving the logical 
quantum information, one can implement all  operations required for universal computation by varying these parity-checking measurements.  
The full process is complex and involves magic state purification \cite{PhysRevA.71.022316,YingMagicState}.

There are different ways in which one can carry out a stabilizer measurement on qubits in separate traps that share some entangled states. We consider two methods whose circuit diagrams are shown in Figure~\ref{fig:ancilla_stabilizer_circuit} (a) and (b). Method (a) uses five nodes, each of which is a module of the kind described in Section~\ref{sec:layout}. The circuit shown in Figure  \ref{fig:ancilla_stabilizer_circuit} (a) effectively induces a parity measurement on the application qubits in nodes $A$, $B$, $C$ and $D$. Node $E$ is an ancilla - the application qubit of node $E$ will be measured and that measurement result determines whether the parity of $A$, $B$, $C$ and $D$ is even or odd. 

A second way of inducing a parity measurement follows the approach in Ref.~\cite{Nickerson2013} and is shown in Figure  \ref{fig:ancilla_stabilizer_circuit} (b). This method uses four nodes, each of which contain two envoy qubits and one application qubit. The idea behind this method is to create a shared 4-qubit GHZ state between the four nodes, which can then be used to generate the non-local parity measurement of the application qubits in node $A$, $B$, $C$ and $D$. The advantages of method (b) over method (a) are that method (b) uses only four nodes, that the operations can be performed parallel and (as we will see) it has higher error correction thresholds. The disadvantage of method (b) is that it requires an extra ion per trap to store an additional envoy qubit and would involve a longer purification protocol. Thus this is not strictly compatible with the ion trap layout shown earlier; we would need to introduce an additional Ca ion and recompile the purification process to this different layout; however, the purification process (denoted by the linked double star symbol in Fig.~\ref{fig:ancilla_stabilizer_circuit}) would remain the same, therefore we can include this case for comparison without generating an explicit low-level blueprint.

The fault-tolerance threshold is an important characteristic of a quantum error correcting code. The concept here is that applying the error-correcting process will actually make things worse, i.e. accelerate the degradation of the logical qubit, if the process of parity measurement is too noisy. Then enlarging the code will actually increase the logical error rate. However, if in fact we find that enlarging the code suppresses errors on the logical qubit, then indeed we are successfully operating within the fault-tolerant threshold. We will set the local gate fidelities to a constant level (corresponding to state of the art numbers) and then consider different levels of network noise, in order to identify the threshold. 

\begin{figure*}
\includegraphics[scale=0.4]{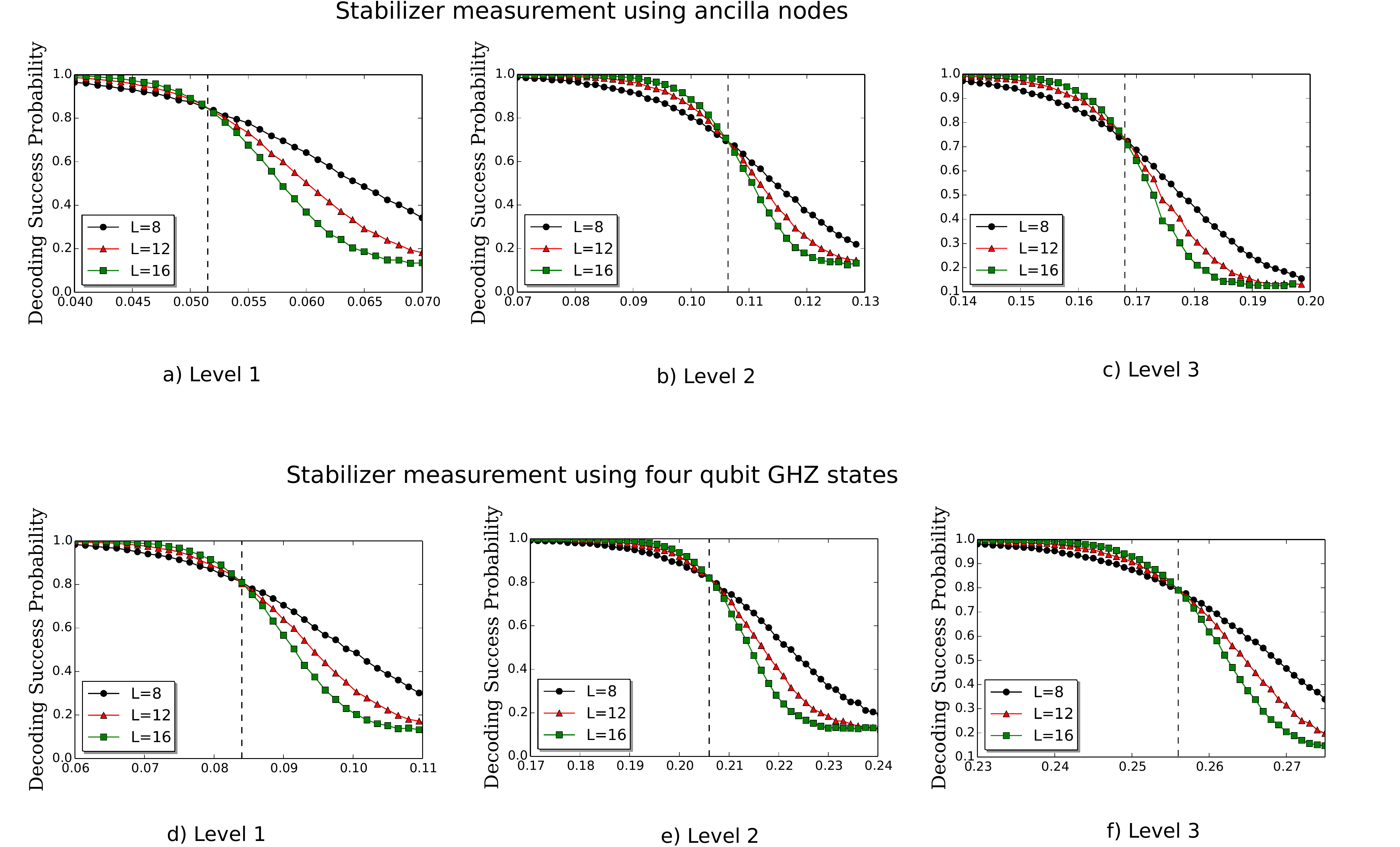}
\caption{(Top row) Results of the threshold calculations for a system using ancilla node to implement the stabilizer measurements (Figure \ref{fig:ancilla_stabilizer_circuit}a)). The logical error rate is calculated as a function of the network error rate $\epsilon$ for the nodes using a) Level 1 b) Level 2 and c) Level 3 purification protocols. (Bottom row) Results of the threshold calculations for a system using GHZ state to implement the stabilizer measurements (Figure \ref{fig:ancilla_stabilizer_circuit}b)). The logical error rate is again found for d) Level 1 e) Level 2 and f) Level 3 purification protocols. The error rates of internal operations are $p_{1}=1\times10^{-6}$, $p_{2}=0.001$ and $p_{m}=0.0005$. The three curves denote results for increasing lattice sizes, where $L = 8$, $12$ and $16$. The threshold is defined as the intersection of these curves which is approximately a) $\epsilon=5.15 \%$ b) $\epsilon=10.64 \%$ c) $\epsilon=16.8 \%$, d) $\epsilon=8.4 \%$ e) $\epsilon=20.6 \%$ and f) $\epsilon=25.6 \%$. The results are obtained 
by averaging 16000 simulation runs. The number of stabilizer measurements after which the decoder attempts to
correct the errors is taken to be $t=4L$, where $2L^2$ is the number of application qubits in the lattice.   \label{fig:threshold}}
\end{figure*}

The thresholds are calculated using Monte-Carlo simulations \cite{PhysRevA.86.032324}.  The procedure is the same that that described in Refs.~\cite{PhysRevX.4.041041,OGorman2016} and indeed the same base numerical code was employed (please see ``/naominickerson/fault\_tolerance\_simulations/releases'' on github.com for the base code). To summarise the procedure: For given local error rates  ($p_{1}=1\times10^{-6}$, $p_{2}=0.001$ and $p_{m}=0.0005$) we select a network size characterized by parameter $L$ such that there are $2L^2$ cells in the toric network. The stabilizer measurement cycles are then simulated. Each stabilizer measurement may introduce error(s), which in turn may induce changes in the stabilizer outcomes in the next cycle. This syndrome information is recorded over $4L$ cycles and then Edmonds' minimum weight perfect matching algorithm is used to attempt to infer the appropriate corrective operations that would recover the ideal state. Consequently the logical qubit either does, or does not, receive an error. This numerical experiment is repeated many times (typically $\sim16,000$) to find the probability that the logically encoded qubit avoids an error, and this is plotted as the $y$-axis of the graphs in Fig.~\ref{fig:threshold}. The entire process is then repeated for a larger $L$, to establish whether this raises or lowers the probability of logial error.  

The procedure outlined above requires as input a set of error rates. For example there will be a specific probability that the correlated error $XYII$ will occur on the four data-qubits involved in a stabilizer measurement; similarly there are specific probabilities for all other error combinations including measurement error on the ancilla qubits. These are pre-calculated by finding the superoperator that describes the effect of the measurement protocol on the input qubits. The superoperator can be obtained by simulating the circuit and making use of the Choi-Jamilkowsky isomorphism. The superoperator can be written as 

\begin{equation}
 \mathcal{S}(\rho)= \sum_{i=0} p_i K_i \rho K^\dagger_i. 
\end{equation}

The map $\mathcal{S}(\rho)$ describes the operation as Kraus operators $K_i$ applied to the input state $\rho$
with probabilities $p_i$, which depend on the chosen protocol, noise model and the error rates. 
The leading term $i=0$ will have corresponding $K_0$ representing the reported parity projection,
and large $p_0$. For the protocols considered here, the other Kraus operations can be decomposed and 
expressed as a parity projection with additional erroneous operations applied. All of $K_i$ can be expressed
as one of the ideal parity projectors followed by (one or more) single qubit Pauli errors. This decomposition then involves two
distinct types of error: `lies', where an incorrect outcome is recorded, and qubit errors, where a physical
error occurs on an application qubit. The probability of each combination of events can be calculated from the values of the $p_i$.
This information on stabilizer performance then enables classical simulation of a full toric code array.

Figure \ref{fig:threshold} shows the results of the toric code simulations using methods (a) and (b) of measuring stabilizers with modules containing Level 1, Level 2 and Level 3 purification protocols. We see that the effect of the purification process has been to tolerate network noise at a very high level; for the five-ions-per-module approach the threshold is $17\%$ noise, and the addition of one further ion can boost this to $>25\%$ noise through the GHZ protocol.

In order to achieve the highest possible thresholds, we exploited the fact that our Level 3 purification results in most of the error probability being associated with a specific one of the three incorrect Bell states (see Eqn.~(\ref{level3channels})). Single qubit gates suffice to move this probability to whichever erroneous Bell state we wish; we moved it to that Bell state which, when employed in the remote gating process Fig.~\ref{fig:ancilla_stabilizer_circuit}(a), gives rise to a $Z$ errror on the ancilla but no error on the data qubit. Ultimately this leads to an incorrect stabiliser result being recorded when the ancilla is measured; this pure `lie' is the most well tolerated type of error in the surface code approach.

Finally we briefly return to the idea of exchanging the roles of the application qubit and the envoy qubit, which we alluded to in Section~\ref{sec:designConsiderations}. In the present context the application qubit is a single ``data qubit'' of the surface code.  The reason to exchange roles is to limit the impact of leakage errors, i.e. errors where an ion leaves the qubit subspace and enters some other state that is not computationally meaningful. Such events are tolerated by the surface code (without explicitly identifying the errors) only if the data qubit is subsequently returned to the computational subspace, so that only a very small proportion of data qubits are in an invalid state at a given time. Since the act of measuring an ion will return it to the correct subspace, an elegant solution is to measure out the ion bearing the data qubit when it is entangled with the envoy, thus teleporting the data qubit onto the envoy which now {\em becomes} the new data qubit. This ``passing of the torch'' means that no single ion will remain unmeasured for very long; however it does leave the new application qubit in the incorrect physical position. Thus one is motivated to apply physical permutation prior to the exchange. Such a permutation also has the benefit that the `old' data qubit can be shuttled to the `rowdy end' of the trap prior to measurement, maintaining the principle that measurement always occurs at that end.

The analysis presented in this paper accounts for the gate and measurement errors, without evaluating the effect of the environmentally induced decoherence on the application qubits. Effectively we are assuming that the raw entanglement rate is sufficiently fast that in the time that it takes to produce a purified envoy qubit, the application qubits accumulate negligible memory errors. Trapped ion qubits can have long decoherence times; recent experiments report a dephasing time $T_2=50$~s and negligible spontaneous decay rates \cite{PhysRevLett.113.220501}. If we assume a 50 second dephasing time for the application qubit, then the time it takes for the fidelity of the application qubit to drop from $1$ to $0.99$ is $\sim 0.725$ seconds. The Level 3 purification protocol requires on average $8$ raw entangled pairs (assuming 10\% network noise). For the decoherence to be negligible we must have $8T_0 \ll 0.725 $ s i.e. $T_0 \ll 0.09$ s, where $T_0$ is the time required to generate one raw entangled pair. Thus the entanglement rate should be significantly greater than 11 Hz. Currently, ion trap experiments report an entanglement rate below $10$~Hz~\cite{Hucul2015}, but in the near future technological improvements promise far higher entanglement rates. As noted earlier this is clearly very desirable in order to achieve high communication rates or fast computer clock speeds. It is also possible that the application qubit's decoherence rate can be reduced either technical improvements or by encoding that qubit over two ions, $\ket{0}\rightarrow\ket{01}$, $\ket{1}\rightarrow\ket{10}$, so as to negate collective phase noise.

\section{Conclusion \label{sec:conclusions}}

We have designed a simple ion trap device for general use as a building block of optically-linked quantum technologies. The unit is capable of interfacing with other similar units over a noisy optical channel, and purifying that channel to enable high fidelity quantum operations between the units. We simultaneously designed a novel purification protocol alongside the device layout in order to meet a number of desiderata for the system. Notably, the proposed system has only five ions in a linear arrangement. These ions (which are of two species) suffice for all entanglement generation, processing, storage and cooling operations. Laser control systems need not differentiate a given ion from its neighbour. 

We evaluated this device in the context of communication over a network of untrusted nodes. Using gate fidelities already reported in the literature we found that each twelve-fold increase in range between the parties leads to a reduction in the rate of communication by a modest factor of $4.8$. This shows the efficiency of the five-ion purification process, however we noted that additional peripheral ions would be desirable as a `buffered' memory and for rapid sequential entanglement attempts.

The same five-ion device was assessed as a component for scalable fault-tolerant computing. Again using reported gate fidelities, we concluded that very high noise can be tolerated in the links of such a system: the threshold was $17\%$. In contrast to the communication scenario, here the five-ion core device will suffice as a building block without additional memory or entangling qubits (although permitting one additional ion will raise the noise threshold further, to $>24\%$).

We conclude that this relatively simple system is a powerful and general device, suitable to be the core of a universal communications node or the building block of a scalable computer. For both applications a key challenge is to increase the rate of entanglement between modules. 

\bigskip

\section{Acknowledgements}

The authors are grateful for helpful discussions with Tom Harty and David Lucas. The authors acknowledge support from the EPSRC National Quantum Technology Hub in
Networked Quantum Information Processing.
\section{Appendix}
\label{sec:Appendix I}
As noted in the main text, there is a simple three-step process by which Alice and Bob, having previously created a high fidelity Bell pair shared between them, can consume this pair in order to perform a cPhase gate between their ``application qubits''. First they each perform a cPhase gate between their Bell qubit and their application qubit. The result can be written
\[
\frac{1}{\sqrt{2}}\left( \ket{00} \mathbb{I}+\ket{11}Z_AZ_B\right)\ket{S}
\]
where $Z_A$ is a single-qubit phase gate ${\rm diag}\{1,-1\}$ acting on Alice's application qubit, $Z_B$ is analogous, and $\ket{S}$ represents both parties' application qubits as well as any other qubits entangled with them, i.e. the ``rest of the system''.

Now Alice measures her Bell qubit in the $x$-basis and Bob does the same but in the $y$-basis. Each receives either a $+1$ or $-1$ on their measuring device, resulting in 
\begin{eqnarray}
&\ &\frac{1}{\sqrt{2}}\bra{X\pm_A}\bra{Y\pm_B}\left( \ket{00}\mathbb{I}+\ket{11}Z_AZ_B\right)\ket{S} \nonumber \\
&=& \frac{1}{2\sqrt{2}}\left(\mathbb{I}+i(\pm1_A)(\pm 1_B)Z_AZ_B\right)\ket{S} \nonumber \\
&=& G\ket{S}\ \ \ {\rm for}\ ++\ {\rm or}\ --,\nonumber \\
&\ &G^\dagger\ket{S}\ \ {\rm otherwise.} \nonumber
\end{eqnarray}
Here $G$ is a diagonal matrix with elements $\{ 1,-i,-i,1 \}$, and we are neglecting global phases $\exp(\pm i \pi/4)$. Matrix $G$ is transformed to the desired cPhase operation, i.e. ${\rm diag}\{ 1,1,1,-1 \}$, by local gates $S={\rm diag}\{ 1,i \}$ performed by both Alice and Bob. Meanwhile $G^\dagger$ instead requires them to apply $S^\dagger$.

\bibliographystyle{unsrt}
\bibliography{ref}

\end{document}